# Role of Multifidelity Data in Sequential Active Learning Materials Discovery Campaigns: Case Study of Electronic Bandgap

**Authors:** Ryan Jacobs,[1,*] Philip E. Goins,[2] Dane Morgan[1,*]

[1] Department of Materials Science and Engineering, University of Wisconsin-Madison, Madison, WI, 53706, United States of America

[2] U.S. C.C.D.C. Army Research Laboratory, 6300 Rodman Road, Aberdeen Proving Ground, MD 21005, United States of America

*Corresponding author e-mail: rjacobs3@wisc.edu, ddmorgan@wisc.edu

## Abstract

Materials discovery and design typically proceeds through iterative evaluation (both experimental and computational) to obtain data, generally targeting improvement of one or more properties under one or more constraints (e.g., time or budget). However, there can be great variation in the quality and cost of different data, and when they are mixed together in what we here call multifidelity data the optimal approaches to their utilization are not established. It is therefore important to develop strategies to acquire and use multifidelity data to realize the most efficient iterative materials exploration. In this work, we assess the impact of using multifidelity data through mock demonstration of designing solar cell materials, using the electronic bandgap as the target property. We propose a new approach of using multifidelity data through leveraging machine learning models of both low- and high-fidelity data, where using predicted low-fidelity data as an input feature in the high-fidelity model can improve the impact of a multifidelity data approach. We show how tradeoffs of low- versus high-fidelity measurement cost and acquisition can impact the materials discovery process, and find that the use of multifidelity data has maximal impact on the materials discovery campaign when approximately five low-fidelity measurements per high-fidelity measurement are performed, and when the cost of low-fidelity measurements is approximately 5% or less than that of high-fidelity measurements. This work provides practical guidance and useful qualitative measures for improving materials discovery campaigns that involve multifidelity data.

## 1. Introduction

Materials play a central role in nearly every aspect of modern society, and the discovery and design of new materials is essential for developing and advancing a multitude of technologies. Despite the importance of developing new materials, historically the commercial deployment of new materials has been slow, and it can take 20 years or longer from initial discovery of a new material to its eventual commercial use.[1] It is worth noting this estimation does not take into account the initial time required to discover a new material, which itself is a time and labor-intensive process. The introduction of the Materials Genome Initiative (MGI) in 2011 sought to significantly reduce the time (and cost) of discovery and deployment of new materials by leveraging the complementary roles of theory, experiment, and computation.[2] In the past 30 years, the methodology of discovering new materials has undergone a pronounced shift from using traditional trial-and-error approaches primarily relying on theoretical and empirical science toward coupled use of computational simulation-based evaluation of new materials with experimental evaluation, and in particular with high-throughput computational exploration and experimental synthesis and characterization.[2,3] In particular, in the past five years, progressively more advanced forms of data-driven approaches leveraging artificial intelligence (AI) and machine learning (ML) have been deployed in the materials discovery and design workflow.[4–12]

Among different methods of materials discovery and design so-called "closed-loop" approaches have recently gained in popularity.[23–27] Closed-loop approaches can generally take different forms, but here we focus on those involving discovery of new materials with desired properties. A closed-loop materials discovery campaign is comprised of numerous evaluation loops, where each evaluation loop consists of a set of experiments, simulations, or some combination of the two to evaluate a set of candidate materials, followed by a data analysis step that suggests, for the next iteration, a new set of materials to evaluate. This data analysis step consists of an acquisition function used to suggest which materials should be evaluated next to meet a specific objective with as few evaluation loops as possible. It is common to employ active learning approaches for the acquisition function's decision making, where the acquisition function uses the machine learning model predictions and, if applicable, uncertainty estimates, to suggest a new set of materials for the next iteration of the closed loop discovery campaign. A materials discovery campaign aided by active learning is sometimes referred to as sequential learning,[28–30] as the ML model used for the active learning continually improves over the duration of the discovery campaign as additional data is included in the model fit. There are many examples of active learning used for computational- and experimental-based discovery campaigns, including developing improved halide perovskite synthesis[31] and crystallization,[32] discovery of new Heusler materials,[33] new iridium oxide polymorphs for oxygen evolution catalysis,[34] discovery of perovskite materials with high piezoelectric coefficient,[35] discovery of materials with high elastic constants[36] and discovery of high entropy alloys with ultra-high hardness,[37] among others. A natural extension of closed-loop materials discovery campaigns is to remove (most) human involvement to realize an autonomous materials discovery system in the form of self-driving laboratories (SDLs). The use of SDLs has exploded recently, with a number of detailed reviews[38–45] and ground-breaking studies[46–49] in just the past couple of years. Overall, the development and use of SDLs is very promising, where Kavalsky et al. estimated the use of SDLs may speed up the materials discovery process by a factor 10-20×,[50] potentially reducing the materials discovery-to-commercialization pipeline significantly.

A important unsolved challenge is determining effective closed-loop materials discovery approaches for iterating on multifidelity data, i.e., data that contains low-fidelity (faster, easier, cheaper) and high-fidelity (slower, more difficult, expensive) methods of evaluating material performance. Important considerations for optimizing sequential learning campaigns with multifidelity data include: how best to encode data of varying fidelity levels in the machine learning models utilized in closed-loop discovery campaigns, how to understand practical tradeoffs of balancing relative data acquisition of new low- vs. high-fidelity data, the role of different relative costs of performing new low- vs. high-fidelity measurements, and the impact of operating on a fixed total budget in order to maximize the efficiency of discovering new materials. It is also worth noting that the impact of using multifidelity data also hinges on relationships between the low- and high-fidelity data, e.g., the strength of correlation between low- and high-fidelity data, and the probability of success of finding a material of interest from the high-fidelity search space. The correlation strength between low- and high-fidelity data will impact the advantage of using multifidelity vs. single fidelity data and optimum data acquisition ratio for a given cost ratio and total budget. For example, a modest correlation between low- and high-fidelity data will likely result in a multifidelity approach being useful, while the limiting cases of no correlation and perfect correlation will result in the best approach being to conduct single fidelity searches over only high-fidelity and only low-fidelity data, respectively. The present work seeks to provide qualitative guidance and attach some concrete numbers to assess each of the key factors raised above, providing a way of thinking about key approaches and tradeoffs for using multifidelity data in closed-loop materials discovery campaigns.

With the rise of SDLs, numerous software packages to simulate and conduct sequential active learning materials discovery campaigns have also been developed. Most relevant for this work, in 2020, Montoya et al. introduced the computational autonomy for materials discovery (CAMD) software package, an open-source python-based framework for sequential learning.[24] In this initial work, Montoya et al. used CAMD, coupled with automated DFT calculations and convex hull phase stability calculations, to discover new stable structures of materials spanning 16 different chemistries. Their work resulted in DFT predictions of hundreds of new, potentially stable binary and ternary compounds in chemical spaces such as Mn-S, Fe-V, V-O, etc. Recently, a follow-on study by Montoya et al.[25] focusing on ternary oxide systems resulted in newly discovered and experimentally confirmed phases of the calcium ruthenate system, further demonstrating the promise of such automated sequential learning approaches for materials discovery. The initial formulation of the CAMD framework used an ML model which conducted active learning using a single fidelity dataset. Follow-on work from Palizhati et al.[29] extended CAMD to also include new methods for handling sequential active learning campaigns with multifidelity data. In the work of Palizhati et al., data fidelity levels were included in the ML model fits via one-hot encoding (e.g., low-fidelity = 0, high-fidelity = 1). A single ML model was used, where material structures may be encoded with elemental or structure-based features, and the data fidelity differentiated by this one-hot encoding. They performed a case study examination using electronic bandgap (similar to this work, but with differences discussed in **Section 2**), and implemented two types of acquisition functions: a Gaussian process regression (GPR) model which can balance exploitation primarily in the high-fidelity predictions and exploration in the low-fidelity candidate space, and a greedy approach which can leverage any ML model and relies on exploitation to suggest new low- and high-fidelity candidates. While they found the multifidelity GPR model performed similarly to a single fidelity support vector regression (SVR) model, their multifidelity SVR model using the greedy approach showed modest discovery improvement compared to the single fidelity SVR model, indicating the use of multifidelity data resulted in a more efficient materials discovery campaign.

Inspired by recent work from Palizhati et al.,[29] we use the CAMD software package and propose a new multifidelity data approach for leveraging low- and high-fidelity data by separately fitting ML models to the low- and high-fidelity data subsets and using the ML model-predicted low-fidelity predictions as an input feature to the high-fidelity ML model. Including the ML-predicted low-fidelity values as a high-fidelity ML model feature allows us to directly exploit the correlation between the low and high-fidelity bandgap values in the sequential learning campaign. The hypothesis we wish to test here is that including a prediction of the low-fidelity value in the high-fidelity model will reduce its error, thus enabling a more efficient sequential learning campaign using multifidelity data. We use our new approach to explore benefits and tradeoffs of using single vs. multifidelity data to conduct materials discovery campaigns with variable cost of low- vs. high-fidelity data and a fixed acquisition budget (i.e., total cost). We use multifidelity electronic bandgap data to provide a demonstration of our approach for the mock discovery of single junction solar cell materials, where we are interested in using our multifidelity data approach to efficiently discover materials with a high-fidelity bandgap value in the range of 1.1-1.7 eV (note the optimum single junction solar cell material bandgap is 1.4 eV).[51] Overall, we find that our new approach of leveraging separate ML models for low- and high-fidelity data results in multifidelity discovery campaigns which outperform previous methods where the data fidelity is distinguished with one-hot encoding. Further, by considering variable cost and a fixed total budget, we find the greatest benefit of using a multifidelity approach is realized when the data acquisition ratio of low- and high-fidelity data is approximately 5 (i.e., 5 low fidelity measurements per high fidelity measurement) and the low fidelity data is roughly about 5% the cost of a high fidelity measurement, or lower, except in cases of very large total budget.

## 2. Data and Methods

In this work, we use DFT-calculated data of electronic bandgaps at different fidelity levels. We use the bandgap database collected by Li et al.[52] This database consists of experimental bandgaps and DFT-calculated bandgaps at the Perdew-Burke-Ernzerhof (PBE) and hybrid Heyd-Scuseria-Ernzerhof (HSE) fidelity levels collected from the Materials Project,[13] OQMD,[17] AFLOW,[14] and JARVIS[21] databases. When performing DFT, PBE calculations are much faster than HSE calculations (typically in the range of 10-50× faster, depending on other computational settings). Further, PBE is lower fidelity than HSE regarding electronic structure calculations, including the electronic bandgap, where it is well known PBE typically underestimates the bandgap by typically 30% but sometimes as much as 100% (i.e., a material which should have a bandgap is predicted to be metallic). Therefore, we focus on using only the PBE bandgaps as our low-fidelity data, and the HSE bandgaps as our high-fidelity data. This differs from the work of Palizhati et al.,[29] who used PBE as low-fidelity and experimental data as high-fidelity. Our choice for using HSE as high-fidelity was motivated by the larger number of HSE than experimental bandgaps in the database, and greater overlap of the same material being present in both the PBE and HSE datasets, enabling use of a larger dataset for our sequential learning campaigns. **Figure 1** shows histogram distributions of the PBE bandgap values (**Figure 1A**) and HSE bandgap values (**Figure 1B**), while **Figure 1C** shows the relationship between the PBE and HSE bandgap values. The original PBE dataset collected by Li et al. is quite large, with 74,992 entries, and the HSE dataset, due to its much higher computational cost, is significantly smaller and contains 5974 entries. When comparing the materials present in the PBE and HSE datasets, there are 4709 materials which are present in both datasets and form the dataset used throughout this work. From **Figure 1A** and **Figure 1B**, it is clear there are a large number of materials with zero bandgap (i.e., metals) in the database, where the inset histograms show the bandgap distributions when metals are removed from the database. From **Figure 1C**, there is reasonably high correlation between the PBE and HSE values ($R^2$ = 0.84), however, there is significant scatter in this correlation at low bandgap values, including within the bandgap design range of 1.1-1.7 eV targeted in the sequential learning campaigns in this work.

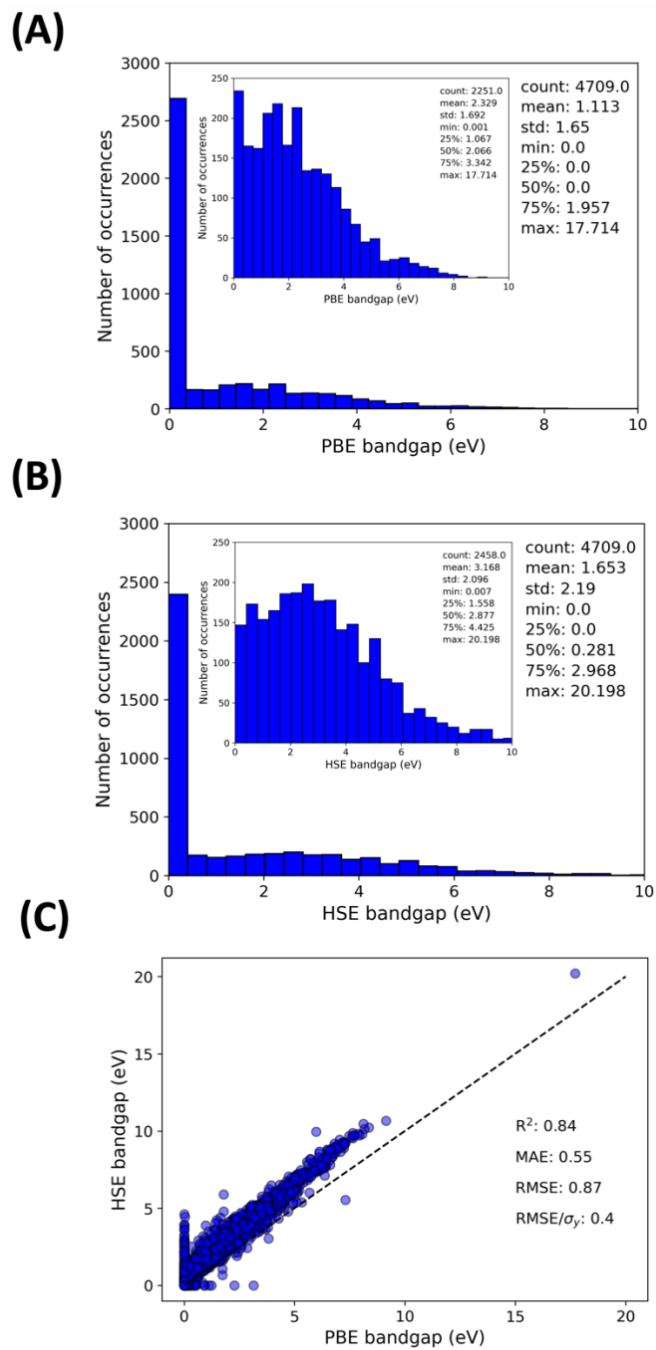

**Figure 1.** (A) Histogram of all PBE (low-fidelity) computed bandgaps. The inset in (A) shows the corresponding histogram of PBE bandgaps after metals (bandgap = 0 eV) are removed. (B) Histogram of all HSE (high-fidelity) computed bandgaps. The inset in (B) shows the corresponding histogram of HSE bandgaps after metals (bandgap = 0 eV) are removed. (C) Parity plot showing relationship between PBE and HSE bandgaps for the subset of 4709 materials for which both PBE and HSE bandgap are available in the database. The provided statistics are for the case where metals are included. If materials with HSE bandgap of 0 are removed, then $R^2$ = 0.67, MAE = 1.05 eV, RMSE = 1.20 eV, and RMSE/$\sigma$ = 0.57.

We extended the CAMD package to incorporate our new method of representing multifidelity data in sequential active learning campaigns. In the *camd.multi_fidelity* module, we added a new *MultiFidelityMLAgent* class, which functions similarly to the *EpsilonGreedyMultiAgent* class from Palizhati et al., but with some key differences. As mentioned above, in Palizhati et al., a single ML model is used and the data fidelity level is distinguished using a one-hot encoding feature in the model fit (e.g., 0 = low-fidelity, 1 = high-fidelity). Here, we use two different ML models: one used to predict low-fidelity data, and one used to predict high-fidelity data. The low-fidelity model uses only trivial-to-calculate elemental property features as input. The high-fidelity model uses these same elemental property features, but also has the ML-predicted low-fidelity values as an additional feature.

Formally, any ML model type can be used to fit the low- and high-fidelity data, and different model types may be used for the low and high-fidelity models. In our specific CAMD implementation, any scikit-learn compatible models can be immediately used by changing a few lines of code. Here, we use a random forest model for both low and high-fidelity models. The feature set was constructed using the MAterials Simulation Toolkit for Machine Learning (MAST-ML) package,[53] and consisted of a standard set of elemental property features.[54–56] Prior to conducting our learning campaigns, we performed some initial fits to our bandgap database, and found a reduced set of 20 features which resulted in robust fits based on random 5-fold cross validation. Doing this initial exploration was necessary to obtain a feature set which was generally reliable (i.e., didn't result in overfitting) and to keep the materials discovery campaigns computationally tractable. In general, our approach can leverage any acquisition function. Here, we used a greedy approach and performed our search using a purely exploitation-based acquisition, where high-fidelity predictions were made and sorted based on their absolute difference from the desired target value of 1.4 eV. A material is considered a promising discovery when its high-fidelity value is in the range of 1.1-1.7 eV, where 260 of the 4709 materials (5.5%) represent the pool of promising materials to be discovered. When conducting a discovery campaign, the user can set the allowed per-iteration budget, and fraction of that budget which should be allocated to high-fidelity measurements. In our approach, the high-fidelity values are queried first, together with the corresponding low-fidelity values for those same materials. If additional low-fidelity measurements are requested (e.g., if the data acquisition ratio is greater than 1, explained more below), then additional low-fidelity values are added, again based on their sorted high-fidelity predictions. For all of our campaigns, we seed the campaigns with the same 50 materials, which we choose randomly as the first 50 materials in our database. If the campaign is a single fidelity campaign, only the high-fidelity data is used. If the campaign is a multifidelity campaign, both the low and high-fidelity values for the same 50 materials are used in the seed. Additional information regarding the CAMD package can be found in the references by Montoya et al.[24] and Palizhati et al.[29]

The goal of the present sequential materials mock discovery campaign is to find the maximum number of new candidate single junction solar cell materials in the bandgap range of 1.1-1.7 eV given a fixed acquisition budget. The performance of the discovery campaign will depend on the relative numbers of low- and high-fidelity data acquired, and their relative cost. As part of this goal, we wish to maximize the multifidelity advantage (MF advantage, discussed below), i.e., use the mixture of low- and high-fidelity data as effectively as possible. Here, we define some key discovery campaign metrics which will be calculated and analyzed throughout **Section 3**. First, the *discovery yield* is the fraction of materials which pass the success criterion (high-fidelity bandgap in range of 1.1-1.7 eV) compared to the total pool of considered materials (here, 4709 different compounds). Note, this is the same discovery yield metric as defined and discussed in the work by Borg et al.[28] The *data acquisition ratio* ($a$) is an input parameter for the campaign, and is the number of low-fidelity measurements performed per high-fidelity measurement in a single loop of the discovery campaign, $a$ = #LF / #HF. The *data cost ratio* ($b$) is the relative cost of performing low-fidelity vs. high-fidelity measurements in a single loop of the discovery campaign, $b$ = cost LF / cost HF, and is used during post-processing to assess campaign performance with respect to cost. The *campaign efficiency* ($c$) is the ratio of areas under the curve (AUC) of the discovery yield of a campaign relative to the discovery yield of an ideal campaign, $c$ = AUC(campaign) / AUC(ideal), and is used to assess campaign performance. An ideal campaign is a campaign where every iteration yields a successful material, and thus the discovery yield increases monotonically as the campaign progresses, and has a value of $c$ = 1 by design. The behavior of the ideal campaign in relation to real discovery campaigns is shown visually and explained in more detail in **Figure 2** and **Section 3.1**. The total cost incurred at any point in the discovery campaign is related directly to the data acquisition and cost ratios, where total cost = 1 + $ab$, in units of number of high-fidelity experiments. Finally, we define the *multifidelity (MF) advantage* as the difference in campaign efficiency $c$ of a campaign using multifidelity data vs. a campaign which uses only single fidelity data, *MF advantage* = $c$(multifidelity campaign) – $c$(single fidelity campaign). The magnitude of the MF advantage will vary based on the data cost ratio and total budget, and maximizing the MF advantage is desired to achieve the most efficient discovery campaign. Note that MF advantage can be negative, which means that the use of multifidelity data results in a less efficient discovery campaign compared to just using data of a single fidelity.

## 3. Results and Discussion

### 3.1. Discovery campaign performance of limiting single and multifidelity cases

In this section, we first analyze the campaign performance of two limiting cases. The first limiting case does not leverage multifidelity data, and instead is a single fidelity campaign searching only over the high-fidelity data. For this single fidelity campaign, the data acquisition ratio has a value of $a = 0$ as no low-fidelity data is queried. The second limiting case uses multifidelity data, but in the limit where all of the low-fidelity data is known *a priori* and can thus be considered prior knowledge. This means that the high-fidelity ML model has access to the exact values of the low-fidelity data as input, and can be considered the ideal, best-case scenario of a multifidelity discovery campaign, at least using the present approach (henceforth, we will refer to this particular campaign as the "ideal multifidelity campaign"). For this ideal multifidelity campaign, all of the low-fidelity data is known *a priori*, so in this case $a = 4709$. **Figure 2** contains discovery campaign plots where **Figure 2A** examines the campaign performance as a function of experiment iteration, ignoring cost, and **Figure 2B** shifts the x-axis to instead be total cost, where cost is denoted in units of high-fidelity measurements (1 HF measurement = 1 total cost unit). For the purposes of the analysis in **Figure 2**, the cost ratio was set to $b = 0.1$. We note this choice of $b$ is arbitrary and case-dependent, and in **Section 3.2** we examine the impact of $b$ on the campaign performance. The choice of $b=0.1$ was motivated mostly by it providing simple benchmark value that is easy to remember and scale. This value implies that HSE is 10× cost of PBE, which is qualitatively reasonable, although in most calculations the HSE is more like 20-50× the cost of PBE. In **Figure 2A**, we observe that the single fidelity campaign (blue curve) with high-fidelity data has a campaign efficiency of $c$ = 0.58. Note that an ideal campaign (dashed line in **Figure 2**) has a value of $c$ = 1 by definition. In addition, we also include a random discovery campaign, which has $c$ = 0.24. The materials discovery campaign using single fidelity data is therefore much better than simple random guessing. More interesting is the impact of multifidelity data in the ideal multifidelity campaign (green curve). Here, the campaign efficiency $c$ = 0.87 is significantly better than the single fidelity campaign. This result demonstrates that, at least under this limiting case condition, our approach of using multifidelity data can significantly enhance sequential learning materials discovery campaigns.

The analysis in **Figure 2A** neglects the impact of cost, and the cost difference between performing low vs. high-fidelity measurements. A more realistic comparison of the campaign performance is shown in **Figure 2B**, where here the x-axis is total cost in high-fidelity units. The single fidelity campaign has slightly reduced campaign efficiency from 0.58 to 0.56 compared to **Figure 2A**. This slight reduction is due to the initial incurred cost of performing 50 high-fidelity measurements and 50 low-fidelity measurements (50 + 50×0.1 = 55 HF cost units) to form the data seed to begin the discovery campaign. By comparison, the campaign efficiency of the ideal multifidelity campaign has a substantial shift from 0.87 down to 0.59, driven by the large initial incurred cost of 50 + 4709×0.1 = 520.9 HF cost units. In addition, the discovery yield of the ideal multifidelity campaign in **Figure 2B** lies below that of the single fidelity campaign for roughly the first third of the discovery campaign, which effect is due to the initial incurred cost effectively pushing the green MF campaign curve in **Figure 2B** to the right, and is therefore a function of the cost ratio $b$. This result indicates that both total cost budget and the cost ratio $b$ play a key role in setting the magnitude of the MF advantage (in this example, MF advantage = 0.59-0.56 = 0.03), in turn affecting decision making of the relative benefits of low vs. high-fidelity data acquisition. The interplay of total budget and cost ratio is discussed in **Section 3.2**, while the effect of data acquisition ratio is discussed in **Section 3.3**.

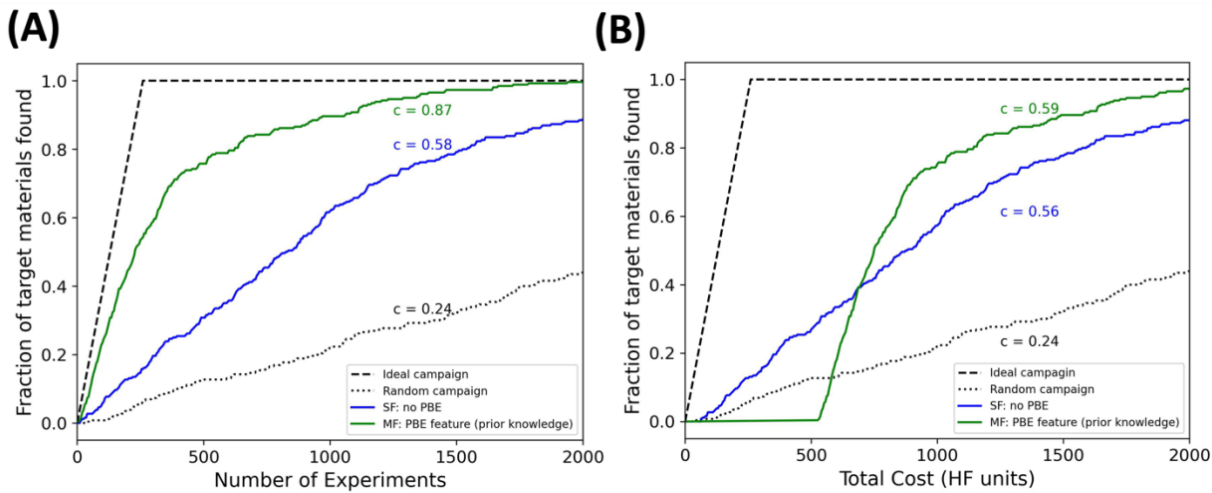

**Figure 2.** (A) Discovery campaign plotted in terms of number of experiments performed. (B) Discovery campaign plotted in terms of total cost (in high-fidelity cost units), assuming data cost ratio $b$ = 0.1. In both plots, the black dash and black dotted lines indicate the performance of an ideal and random campaign, respectively. The blue lines are the performance of a single fidelity discovery campaign using only high-fidelity data. The green lines are the performance of an ideal multifidelity campaign where all low-fidelity data is known exactly *a priori* and thus used as direct input to the model.

### 3.2. Impact of cost ratio and total budget on campaign performance

In the previous section, our analysis was limited to a cost ratio of $b$ = 0.1 for the single fidelity and ideal multifidelity limiting cases. Here, we perform a more detailed analysis of the role of multifidelity data in enhancing the discovery yield for the same ideal multifidelity data case discussed above (where all low-fidelity data is treated as prior knowledge), where we quantify the MF advantage as a function of cost ratio $b$ and total budget, as shown in **Figure 3**. **Figure 3** is a contour map quantifying the MF advantage as a function of cost ratio $b$ vs. total budget. An immediate observation from **Figure 3** is that, even in this limiting case where all low-fidelity data is known, thus amplifying the benefit of using multifidelity data, only a limited domain of *b-budget* space shows an MF advantage > 0 (the two brightest color regions in **Figure 3**). The benefit of a multifidelity approach is enhanced in the limit of large total budget and modest cost of low-fidelity data (e.g., $b$ < 0.1). For small total budgets (e.g., total budget of 250 HF cost units), a multifidelity approach only makes sense if the low-fidelity data are extremely cheap to obtain, e.g., 1% the cost of a high-fidelity measurement. Contour plots to quantify the MF advantage like those in **Figure 3** are useful to inform decision making of whether to include multifidelity data in the discovery campaign or simply perform sequential learning over the high-fidelity measurements only. We will further examine maximization of the MF advantage as a function of *b* and total budget by also including different data acquisition ratios *a* in **Section 3.4**, but first, we turn to examining the impact of the data acquisition ratio *a* on the discovery yield in **Section 3.3**.

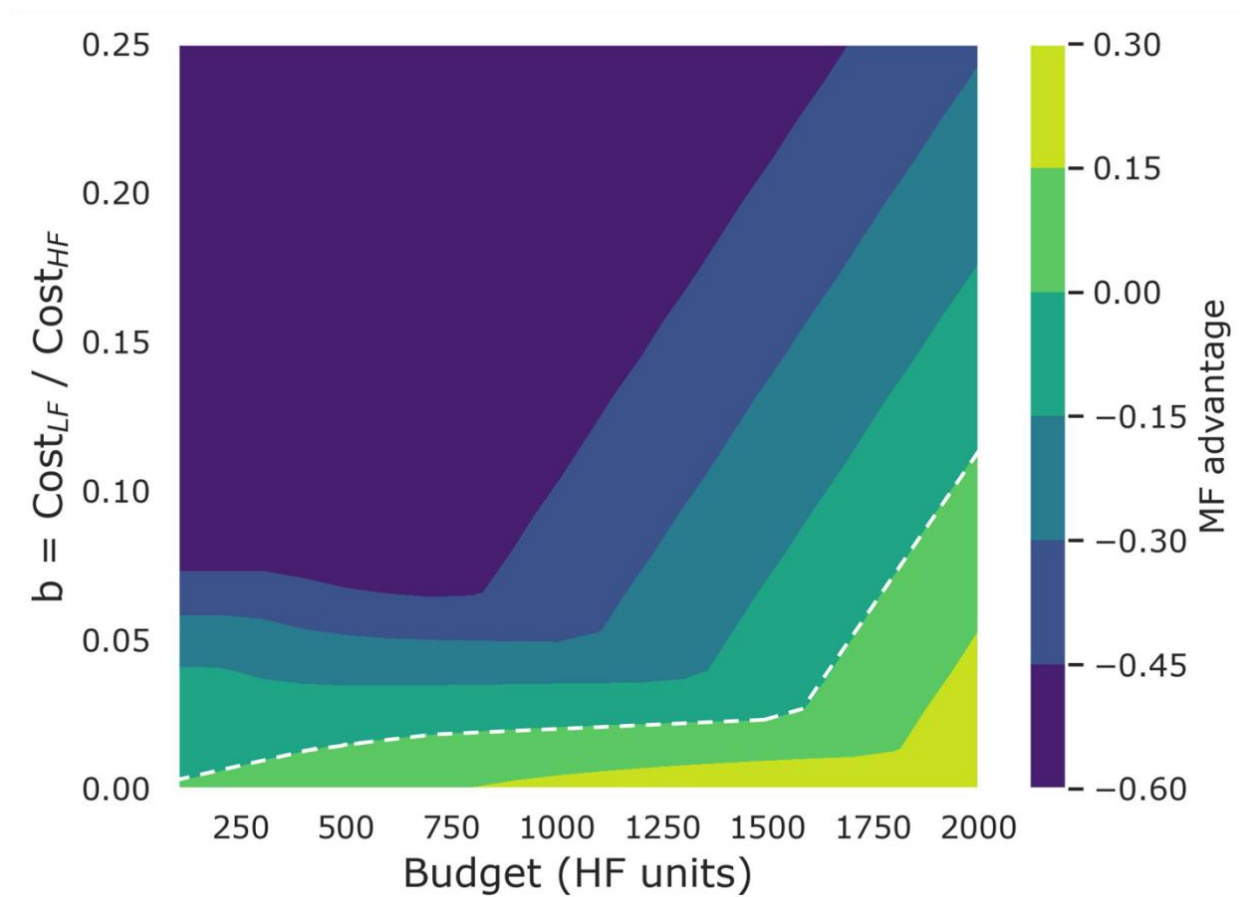

**Figure 3.** Contour map displaying the advantage afforded by using a multifidelity data approach in the materials discovery campaign, for the case of an ideal multifidelity campaign where the low-fidelity data are known exactly *a priori*. The map displays the data cost ratio *b* versus the total budget (in units of high-fidelity measurements), with heats denoting the magnitude of multifidelity advantage, defined as the improvement in discovery campaign efficiency when both high and low-fidelity data are used versus only running a single fidelity discovery campaign. The white dashed line denotes the boundary between negative and positive MF advantage, where regions to the right and below the white dashed line denote MF advantage > 0. This region represents the conditions where a MF approach is beneficial.

## 3.3. Impact of data acquisition ratio on campaign performance

Thus far, our analysis on the impact of multifidelity data on materials discovery campaigns has only considered two limiting cases where the data acquisition ratio $a$ = 0 (single fidelity using only high-fidelity data) and $a$ = 4709 (ideal multifidelity scenario). In practice, a full suite of low-fidelity data may not be available or practical to obtain prior to performing a materials discovery campaign, and must instead by acquired on-the-fly. Here, we examine the impact of varying the data acquisition ratio $a$ on the discovery yield, ignoring the impact of cost and total budget. In **Section 3.4**, we include analysis of tradeoffs between cost ratio and total budget to find an optimum data acquisition ratio which maximizes the benefit of using a multifidelity approach. **Figure 4A** contains discovery yield curves for varying data acquisition ratios using our multifidelity data approach. The ideal, random, and limiting cases investigated in **Section 3.1** are included as well. We observe from **Figure 4A** that the discovery yield improves monotonically when increasing $a$ from 1 to 20, where the campaign efficiencies are $c$ = 0.58, 0.70, 0.76, 0.79, 0.80, and 0.81 for $a$ = 1, 2, 3, 5, 10 and 20, respectively. As a reference, the campaign efficiencies for the single fidelity and ideal multifidelity runs are $c$ = 0.58 and $c$ = 0.87, respectively. The improvement of the discovery campaign with increasing data acquisition ratio stems from the rate of improvement of the low-fidelity data ML model, which, given the use of the low-fidelity prediction as a feature in the high-fidelity ML model, leads to improvement in the high-fidelity predictions, as shown in **Figure 4C** (low-fidelity model) and **Figure 4D** (high-fidelity model). Given that our acquisition function is based on a greedy approach relying purely on exploitation, it makes sense that higher data acquisition ratios produce higher campaign efficiencies by virtue of more accurate predictions of the underlying ML models. In **Figure 4B**, we perform discovery campaigns using the same set of data acquisition ratios, but instead of using our present multifidelity approach leveraging separate low- and high-fidelity data ML models, we use a random forest model but with the greedy multifidelity approach available in the CAMD package and used in work by Palizhati et al., where a single ML model is used and the data fidelities are distinguished by one-hot encoding. In **Figure 4B**, we observe modest improvement of the discovery campaign when increasing $a$ from 1 to 20, where the campaign efficiencies are $c$ = 0.58, 0.62, 0.60, 0.64, 0.68 and 0.67 for $a$ = 1, 2, 3, 5, 10 and 20, respectively. It is noteworthy that our present approach of using two ML models for running multifidelity sequential learning campaigns results in a higher campaign efficiency than the previous approach of Palizhati et al., at least for this dataset.

There are two additional features of **Figure 4A** that are worth discussing. The first is that the multifidelity campaign with $a$ = 1 (purple line) shows no improvement over the single fidelity campaign where no low-fidelity data is used (blue line). This behavior is likely the result of the low-fidelity ML model predictions having too high of an error to noticeably improve the predictions of the high-fidelity ML model and thus aid the discovery process. Instead, the ML-predicted low-fidelity values used as a feature in the high-fidelity ML model, with their high RMSE of > 1 eV, effectively functions as a highly noisy feature with minimal correlation to the high-fidelity bandgap values. While the case of $a$ = 1 shows no improvement from the standpoint of discovery campaigns cast in terms of number of experiments, the performance relative to the single fidelity case will actually become worse when casting the discovery campaign in terms of cost, due to the incurred cost of performing the additional low-fidelity measurement for each high-fidelity measurement, where this cost increase per iteration is equal to $b$ (since $a$ = 1), thus resulting in a reduced campaign efficiency. The second feature worth noting is the observed diminishing returns of discovery yield for higher $a$ values, where $a$ = 10 and $a$ = 20 have essentially the same campaign efficiency ($c$ = 0.80 and $c$ = 0.81, respectively). These diminishing returns are likely due to the inherent error in the low-fidelity ML predictions, where for a model trained on all of the data, the RMSE of the low-fidelity predictions is still about 0.3 eV. The low-fidelity predictions used as a feature in the high-fidelity ML model therefore have a typical RMSE of about 0.3 eV. Our present approach uses the ML-predicted low-fidelity values in the high-fidelity ML model, rather than the exact low-fidelity values (except in the ideal multifidelity case). It is possible the higher data acquisition cases may approach the ideal multifidelity case after fewer experimental iterations if the exact low-fidelity values were used instead of the ML-predicted values. However, using the ML-predicted values in this case enables us to highlight another important aspect present in many practical scenarios, and that is the role of error in measurements. The difference in campaign performance from $a$ = 20 and $a$ = 4709 is likely the result of the 0.3 eV error in the low-fidelity values. While this RMSE of about 0.3 eV is much smaller than the standard deviation of the high-fidelity dataset we are searching over (standard deviation of 2.19 eV, see **Figure 1**), it nonetheless causes a decrease in campaign efficiency from $c$ = 0.87 (ideal multifidelity campaign) to $c$ = 0.81 ($a$ = 20). As discussed above in the context of **Figure 2**, depending on the total budget available, for some cases it may be a detriment to perform a large number of low-fidelity measurements prior to initiating the sequential learning campaign over the high-fidelity compound search space.

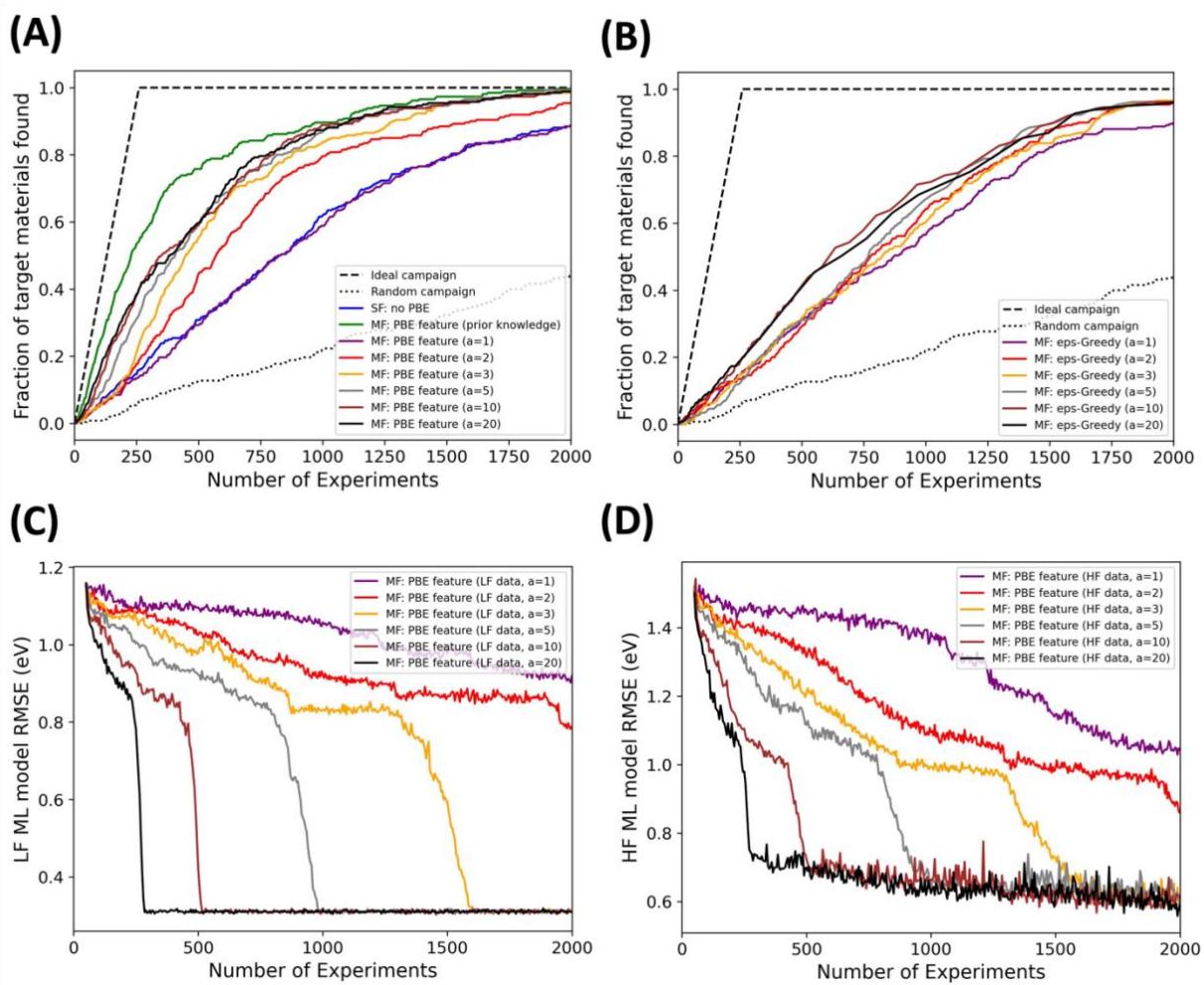

**Figure 4.** (A and B) Discovery yields of multifidelity campaigns with varying data acquisition $a$ ratios: $a$ = 1 (purple), $a$ = 2 (red), $a$ = 3 (orange), $a$ = 5 (grey), $a$ = 10 (brown), $a$ = 20 (black) curves. In (A), the multifidelity approach of the present work is used, in (B) the epsilon-greedy multifidelity agent from the CAMD package is used. In (A), we also include the same single fidelity (blue) and ideal multifidelity using low-fidelity data as prior knowledge (green) from **Figure 2**. In (A), the campaign efficiency values are $c$ = 0.58, 0.70, 0.76, 0.79, 0.80, and 0.81 for $a$ = 1, 2, 3, 5, 10 and 20, respectively. In (B), the campaign efficiency values are $c$ = 0.58, 0.62, 0.60, 0.64, 0.68 and 0.67 for $a$ = 1, 2, 3, 5, 10 and 20, respectively. As a reference, the campaign efficiencies for the single fidelity and ideal multifidelity runs are $c$ = 0.58 and $c$ = 0.87, respectively. Panels (C) and (D) are low-fidelity (panel C) and high-fidelity (panel D) ML model RMSE values on the full dataset as a function of number of experiments, where the colors denoting different acquisition ratios $a$ are the same as in panels (A) and (B).

### 3.4. Maximizing the multifidelity data advantage

The choice of performing sequential materials discovery campaigns using single or multifidelity methods hinges on the MF advantage being positive, meaning the use of multifidelity data yields a higher campaign efficiency than the use of single fidelity data alone. The MF advantage depends on the data acquisition ratio $a$, data cost ratio $b$, and total budget. In practice, knowledge of total budget and cost ratio $b$ should be known at the outset of a materials discovery campaign, and the exercise therefore comes down to choosing the data acquisition ratio such that the MF advantage is maximized. **Figure 5** shows contour plots for varying data acquisition ratios, extending the analysis discussed in **Section 3.2** for the context of the ideal multifidelity campaign to include the same set of data acquisition ratios discussed in **Section 3.3** above. The same general trends from **Figure 3** persist in our interpretation of **Figure 5**, namely, that the MF advantage increases in the limit of large budget and cheap low-fidelity data. Interestingly, the maximum MF advantage of 0.301 can be realized for an intermediate data acquisition ratio, $a$ = 5, meaning that the use of our multifidelity data approach yields a maximum campaign efficiency that is 0.301 higher than the corresponding single fidelity campaign which uses only high-fidelity data. The result that an intermediate data acquisition ratio maximizes campaign efficiency makes sense when considering the tradeoffs of data cost ratio and total budget: if $a$ is too small (e.g., $a$ = 1), the low-fidelity ML model error does not reduce sufficiently fast to yield a high-fidelity ML model with low enough error to result in an improved discovery yield. On the other hand, if $a$ is too large (e.g., $a$ = 20), the benefit of quickly exploring the space of low-fidelity materials does not persist to longer discovery campaigns (i.e., larger total budgets), because the costs incurred early in the campaign become too high to offset the benefit of a more predictive high-fidelity ML model. It is interesting to note, perhaps by coincidence, that in the study by Gantzler et al.,[57] which used multifidelity Bayesian optimization for discovering covalent organic frameworks for gas separations, they found an optimum discovery campaign used 38 low-fidelity and 9 high-fidelity measurements, an acquisition ratio of 38/9 = 4.2, close to our ideal value of 5 discussed above. In their work, the cost ratio (given in terms of calculation time) of low- vs. high-fidelity was 15 minutes vs. 230 minutes, a ratio of 15/230 = 6.5%, again quite similar to our cost ratio cutoff of about 5% denoting when the use of multifidelity data becomes most beneficial. While these similarities are encouraging support for the generality of our results, the nature of correlation between the low-fidelity and high-fidelity model and the accuracy of the ML model will play a significant role in the exact values of this result for both these and other cases. Similarly, the choice of acquisition function and active learning approach may also impact the outcome. However, we expect the qualitative behavior observed in this work to be quite universal due to its dependence on simple logical principles. Since the acquisition function is chosen by the user it would be an interesting subject of future work to assess how it impact the results, as it may be possible to find better exploration approaches for this type of multifidelity problem.

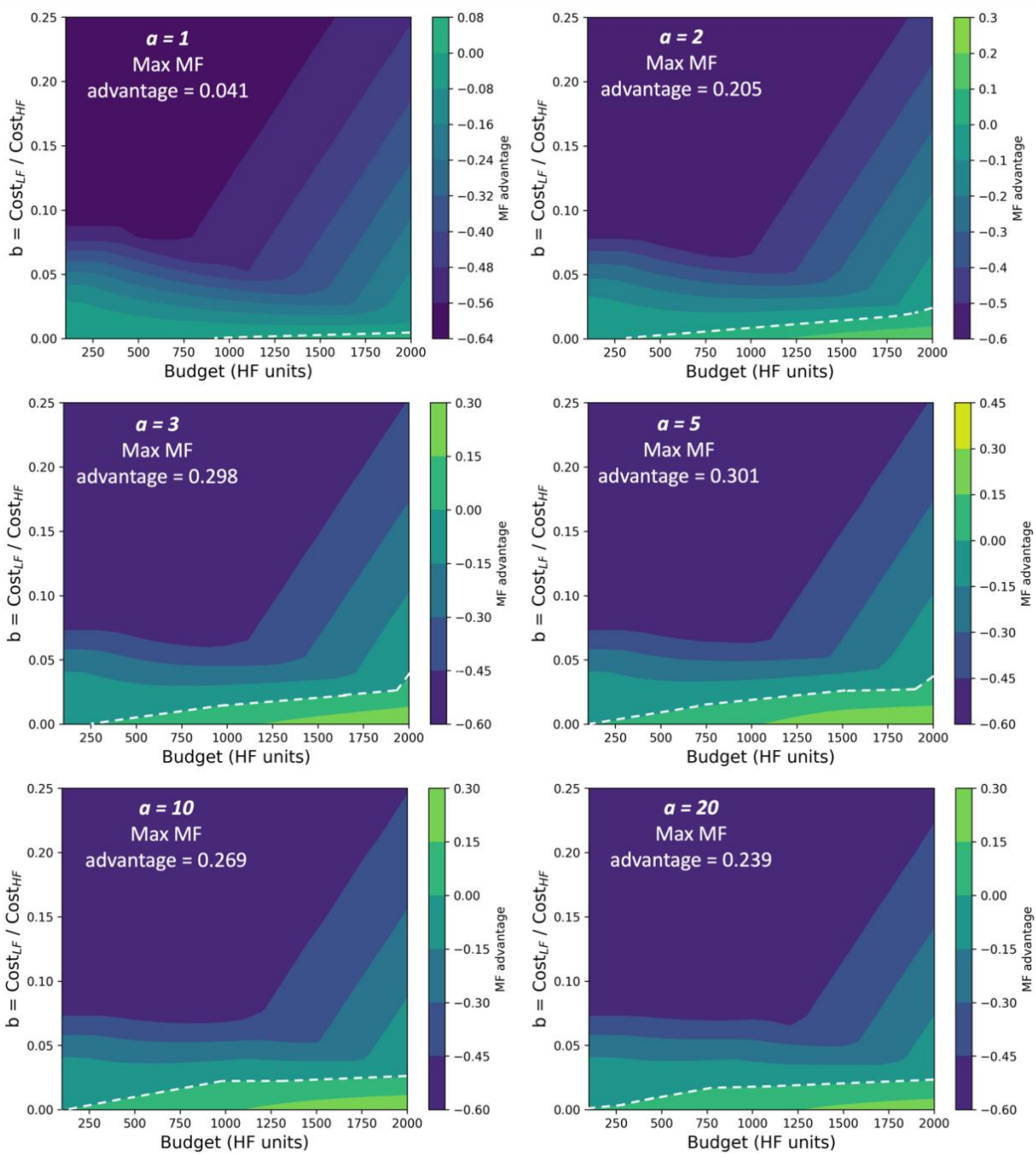

**Figure 5.** Contour map displaying the advantage afforded by using multifidelity data approach in the materials discovery campaign, for the case of varying data acquisition ratio *a* from *a* = 1 to *a* = 20. The map displays the data cost ratio *b* versus the total budget (in units of high-fidelity measurements), with heats denoting the magnitude of MF advantage, defined as the improvement in discovery campaign efficiency when both high and low-fidelity data are used versus only running a single fidelity discovery campaign. For this set of data acquisition ratios, the MF advantage is maximized for *a* = 5. The white dashed lines denote the boundaries between negative and positive MF advantage, where regions to the right and below the white dashed lines denote MF advantage > 0.

### 3.5. Additional Discussion

Our multifidelity approach explored in this work is effective in deciding if, and to what extent, multifidelity data will be effective, and includes an example case study of this decision-making process. When approaching a new closed-loop multifidelity data problem, we believe it is important to have some baseline knowledge of the following aspects. First, how large is the possible design space, and what is the probability of success? That is, what fraction of possible materials may meet the stated design goals? Such knowledge is important to understand the possible success rates of random guessing vs. a sequential active learning approach. Some knowledge of the fraction of possible materials that may meet the design goal enables construction of discovery yield curves. If the goal is just discovery of a single new material, then knowledge of the probability of success is likely more important than an estimation of the total number of materials under consideration in the design space. These above considerations mesh well with some key conclusions drawn by Borg et al.[28] when they examined sequential active learning campaigns with single fidelity data, where they found that the ML model performance for materials discovery depended strongly on the target range of the property distribution which yields success, and whether one is interested in a single discovery or many discoveries. The second aspect to keep in mind when approaching a new problem of this type is the strength of the correlation between low- and high-fidelity data. In the present work, there was significant correlation between low- and high-fidelity bandgap values, and the existence of such correlation resulted in significant benefit of using ML-predicted low-fidelity values to guide prediction of high-fidelity values to discover new materials. In the limit that low-fidelity correlation with high-fidelity becomes weak (e.g., if there is significant uncertainty in the measurements), we anticipate the benefit of a multifidelity approach will reduce, and it is possible a single fidelity search using just high-fidelity data may be more effective. The third aspect to bear in mind is the practical considerations of total budget and cost ratio between low- and high-fidelity measurements. Borg et al.[28] also pointed to the number of iterations in a discovery campaign being a key marker of model performance, where here we would add that for multifidelity campaigns, data cost ratio also plays a key role in this assessment. Provided knowledge of the design space, probability of success, and correlations between low- and high-fidelity data are known to at least some extent, then the approach and analysis conducted here is effective for gauging the anticipated payoff of using multifidelity vs. single fidelity data for sequential materials discovery campaigns, thus providing a means to realize more efficient discovery of new materials.

This work is meant as a case study to demonstrate an approach and way of thinking about assessing cost-benefit tradeoffs to optimize materials discovery under a fixed budget and how to leverage multifidelity datasets most efficiently. As discussed in previous works by Borg et al.[28] and Palizhati et al.,[29] it is difficult to draw universal conclusions about sequential learning campaign behavior because virtually every aspect of the campaign performance depends on the exact dataset and materials design problem being considered. However, we believe that our results showing high MF advantage in the limit of large budgets and low data cost ratio (i.e., cheap low-fidelity data) are likely to be universal, because using a multifidelity approach will always be valuable in the limit of vanishingly cheap low-fidelity data (data rich) and high budget (resource rich) limiting cases. A possible exception to this rule would be a scenario where the probability of success is sufficiently high such that random guessing yields comparable performance to performing a sequential active learning campaign, but we foresee such a scenario as being unlikely to occur in practice.

## 4. Summary and Conclusion

In this work, we implement a new multifidelity data approach for improved sequential active learning materials discovery campaigns. Contrary to previous work which used a single ML model trained on both low- and high-fidelity data, where data fidelity was differentiated by a one-hot encoding feature, here we utilized two separate ML models: one trained only on low-fidelity data, and one trained on high-fidelity data using the ML-predicted low-fidelity values as an input feature. Using our approach, we analyzed the tradeoffs between total budget, data cost ratio, and data acquisition ratio and their resulting impact on campaign discovery yield, and quantified the relative benefit of using multifidelity data over single fidelity data as a function of these variables for mock discovery of new single junction solar cell materials with an electronic bandgap in the range of 1.1-1.7 eV. We find that our new approach of representing multifidelity data modestly out-performed previous approaches, at least for this dataset, indicating that the effectiveness of using separate ML models for low- and high-fidelity data with ML-predicted low-fidelity values as a feature in the high-fidelity model. Further, we find that the optimum data acquisition ratio $a = 5$ resulted in materials discovery campaigns with highest MF advantage of about 0.3, where the highest MF advantages can be realized when the cost of low-fidelity measurements is roughly 5% or less than high-fidelity measurements. We believe that although the quantitative results in this work will depend on specifics of our data and acquisition function, the trends observed in this work are likely to be applicable for many multifidelity materials design campaigns.

**Acknowledgements:** This project was funded by the United States Department of Defense Army Research Laboratory under award number W911NF2120040. The authors thank Joseph Montoya for helpful discussions.

**Conflicts of Interest**

The authors of have no conflicts of interest to declare.

**Data Availability**

The database of computed bandgaps used in this work, along with data files to recreate the figures are available on Figshare: https://doi.org/10.6084/m9.figshare.24429028.v1. The code used in this study is available from the above Figshare link and on Github: https://github.com/uw-cmg/CAMD/tree/main.